\documentclass[preliminary,copyright]{eptcs}
\providecommand{\event}{GRAPHite 2012} 

\usepackage[utf8]{inputenc}
\usepackage[T1]{fontenc}
\usepackage[cmex10]{amsmath}
\usepackage{amsfonts}
\usepackage{amssymb}
\usepackage{algpseudocode}
  \usepackage[pdftex]{graphicx}
\usepackage{comment}

\title{A structural analysis of the A5/1 state transition graph\thanks{
Partially supported by the DFG grant ME 3250/1-3,
and by MADALGO -- Center for Massive Data Algorithmics,
a center of the Danish National Research Foundation.
}}
\def\titlerunning{A structural analysis of the A5/1 state transition graph}

\author{
Andreas Beckmann
\institute{Goethe-Universität Frankfurt\\
Institut für Informatik\\
Frankfurt am Main, Germany}
\email{beckmann@cs.uni-frankfurt.de}
\and
Jaroslaw Fedorowicz
\institute{Goethe-Universität Frankfurt\\
Institut für Informatik\\
Frankfurt am Main, Germany}
\email{fedorow@cs.uni-frankfurt.de}
\and
Jörg Keller
\institute{FernUniversität in Hagen\\
Fakultät für Mathematik und Informatik\\
Hagen, Germany}
\email{joerg.keller@fernuni-hagen.de}
\and
Ulrich Meyer
\institute{Goethe-Universität Frankfurt\\
Institut für Informatik\\
Frankfurt am Main, Germany}
\email{umeyer@cs.uni-frankfurt.de}
}
\def\authorrunning{A. Beckmann, J. Fedorowicz, J. Keller \& U. Meyer}

\begin{document}
\maketitle

\begin{abstract}
We describe efficient algorithms to analyze the cycle structure
 of the graph induced by the state transition function of the A5/1 stream cipher
 used in GSM mobile phones and report on the results of the implementation.
The analysis is performed in five steps utilizing HPC clusters, GPGPU and external memory computation.
A great reduction of this huge state transition graph of $2^{64}$ nodes is achieved by focusing on special nodes in the first step
 and removing leaf nodes that can be detected with limited effort in the second step.
This step does not break the overall structure of the graph and keeps at least one node on every cycle.
In the third step the nodes of the reduced graph are connected by weighted edges.
Since the number of nodes is still huge an efficient bitslice approach is presented that is implemented with NVIDIA's CUDA framework and executed on several GPUs concurrently.
An external memory algorithm based on the STXXL library and its parallel pipelining feature
 further reduces the graph in the fourth step.
The result is a graph containing only cycles that can be further analyzed in internal memory to count the number and size of the cycles.
This full analysis which previously would take months can now be completed within a few days
 and allows to present structural results for the full graph for the first time.
The structure of the A5/1 graph deviates notably from the theoretical results for random mappings.

\end{abstract}


\section{Introduction}

GSM (Global System for Mobile Communications) is the set of standards for cellular networks of the second generation and dates back to 1990.
Even though the fourth generation of mobile communication standards was introduced in 2010,
 more than 75\% of all mobile connections that were established worldwide at the end of 2010 were using GSM~\cite{GSM2010}.
The security of GSM-based communication and the underlying A5 stream cipher family has been studied by
several groups, see e.g. \cite{cryptome:A5, Golic1997, Nohl}.

We present an efficient method to analyze the cycle structure of the directed graph $G=(V,E)$
 induced by the A5/1 stream cipher,
 which serves to encrypt speech data over the air between GSM mobile phones and the base station.
Each state of this stream cipher can be described using 64 bits and
 the set of states corresponds to the nodes $V$.
A state transition function $f: V \mapsto V$ can be derived and
 the edges can be computed as $E = \{ (x, f(x)): x \in V \}$.
Each node of graph $G$ has an outdegree of one, so that $|V| = |E| = 2^{64}$ and the graph consists of one or more weakly connected components (WCC).
Each WCC contains one cycle with root directed trees attached to the cycle and the root node being part of the cycle.
The cycle structure (number and sizes of cycles and WCCs) can be compared to expected values known for random mappings,
e.g.\ $\Theta(\sqrt n)$ nodes are expected to be on cycles in a random mapping of size $n$.
Notable deviations from these expectations, in particular cycles shorter than the already small expectation,
may indicate a weakness of the cipher because these non-random structures might enable various types of attacks.

The size of the graph prevents an explicit construction in internal or external memory,
thus excluding algorithms that modify the original graph structure during a reduction,
and calls for specific algorithmic solutions to reduce the data size to a feasible level.
In contrast to prior work, our approach takes days instead of months by tailoring the hardware and algorithm for each solution step,
 and derives much more information about the graph than only the cycle lengths.

Our new findings complete and confirm previous partial results and estimated properties of the A5/1 graph.
The techniques used for the A5/1 graph may be adjusted to different random mappings, e.g.\ hash chains~\cite{Lamport81a}.



The remainder of this work is organized as follows: in Section \ref{chap:basics} we give a short description of the A5/1 stream cipher in general and derive some important properties.
Related work on this topic is summarized in Section \ref{chap:prev}.
In Section \ref{chap:algo} the five steps to retrieve the structure information of the A5/1 state transition graph are described.
Details on the implementation of the algorithms are presented in Section~\ref{chap:impl}
 and the results are discussed in Section~\ref{chap:results}.
We conclude in Section~\ref{chap:conclusion}.

\section{The A5/1 stream cipher}
\label{chap:basics}
\begin{figure}[ht]
        \centering
        \includegraphics{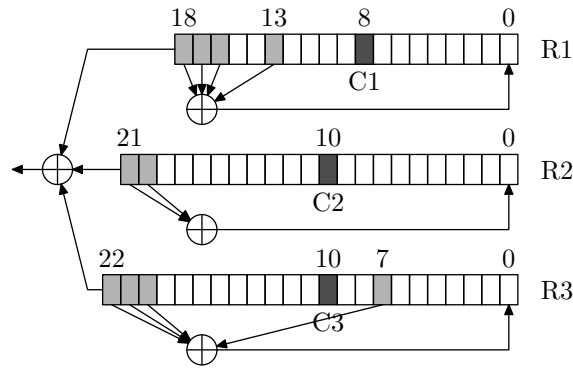}
        \caption{The linear feedback shift registers of the A5/1 stream cipher}
        \label{fig:A51}
\end{figure}
\noindent
In the following we provide a description of the internal state of the A5/1 stream cipher.
Its application for en-/decrypting GSM communication is described in~\cite{Biham2000}.

The A5/1 stream cipher consists of three different irregularly clocked linear feedback shift registers (LFSRs) that are combined via a clock control as shown in Figure \ref{fig:A51}.
Whenever a register is clocked, the feedback bits (e.g.\ 13, 16, 17, and 18 for $R1$) are XORed and inserted into bit 0 after a left shift.
The feedback taps of the three LFSRs in the A5/1 stream cipher were chosen in a way that the registers have maximum length periods,
 i.e.\ all other possible states of a register will be generated before a state will be generated for the second time.
The all-zero value of a register is not allowed as it would not lead to a different state, therefore register $R1$ has $2^{19}-1$, $R2$ has $2^{22}-1$ and $R3$ has $2^{23}-1$ valid and reachable states.

To determine which register is clocked in each iteration of the A5/1 stream cipher, each register has a bit position marked as the clock tap (C1, C2 and C3) and a majority clock function takes these three bits as arguments.
A register is clocked if its clock bit equals the majority value of the three clock bits.
That means that either two or all three registers are clocked at the same time in each iteration.

The values of the three clock bits form eight different combinations.
For each clock bit there are two combinations where this bit differs from the other two causing it not to be clocked.
Therefore a single register is clocked in three out of four cases.

The 64 bit value obtained by concatenating the bits of the three LFSRs describes a single state $x$ of the A5/1 stream cipher.
Clocking the registers according to the majority clocking rule leads to the follow-up state $f(x)$.
Note that the state transition function can be inverted easily so that backward clocking of the stream cipher is possible and produces up to four different predecessors states.


Since the A5/1 graph is based on LFSRs, the following observations have been made.
A cycle in the A5/1 generator implies that all three LFSRs have been cycled at least once.
It is not possible to generate a cycle while keeping one register unclocked.
Therefore it is possible to restrict observation to a fixed value in one register.
Each cycle in the graph will contain at least one state with this fixed register value.
Since following the state transition function from an arbitrary state finally leads to a cycle, it also leads to a state with the fixed value.

The minimum size $L$ of a cycle in the state transition graph can be calculated as follows:
Since the LFSRs in A5/1 stream cipher have maximum length periods, each cycle contains nodes that are built from all valid states of every register, especially the $2^{23}-1$ valid states of the largest register $R3.$
A single register is expected to be clocked in three out of four cases therefore the whole stream cipher must be clocked $L= \frac{4}{3} (2^{23}-1) = 11184809.\overline{3}$ times on average to see all values of $R3$.
The average R3 cycle length $L$ thus bounds the minimum cycle length,
and other cycle lengths are multiples of $L$.
On a path from an arbitrary state to the cycle the fixed register value will appear regularly (every $L$ steps on average).

\section{Related work}
\label{chap:prev}
This work is based on previous work by Beckmann and Keller \cite{Beckmann2007}.
They described the algorithm and method that is able to fully analyze the structure of the A5/1 Graph.

One evaluation of the state transition function of the A5/1 stream cipher maps each state to another state in the same state space.
In the literature these mappings are often called \textit{random mappings}, because it is  assumed that the mapping function maps a state to another random state.
Most random mapping graphs are derived from hash functions, pseudo random number generators (PRNGs) or stream ciphers like the A5/1.
Flajolet and Odlyzko~\cite{Flajolet1990} studied several average case properties of random mapping graphs.

Other expected properties of the graph produced by the A5/1 stream cipher were described by Golic~\cite{Golic1997}.
He analyzed the branching process of the root directed trees that are attached to the cycles by the root of each tree.
The branching process of the trees can be modeled by a Galton-Watson branching process with the expected number of offsprings $\mu=1$ and a related variance of $\sigma^2=\frac{9}{8}$.
A Galton-Watson process with an expected number of one child in each generation is called critical because the probability that the generated tree is limited in size equals one.

The properties of the model above only hold under the assumption that the branching probability of each node in the sub-tree is independent of the branching probability of the nodes in the same or in the preceding levels.
This is not guaranteed in the special case of the A5/1 graph because there is a weak dependency between these probabilities that might influence the expected properties.
We will compare these theoretical values with our observed values in Section \ref{chap:results}.

Keller \cite{Keller2007} provides first details on the number of components and the length of the largest cycle which leads to the result that the graph produced by the A5/1 stream cipher looks non-random.
Later the number of components was further refined in \cite{Beckmann2007}.

Exact results have not been published yet, so that our work is the first to describe the exact number of components together with other details.

\section{The Algorithms}
\label{chap:algo}
To completely analyze all cycles and their properties in the A5/1 state transition graph a five step approach is described in this section.
The first step reduces the state transition graph of size $2^{64}$ to a \emph{candidate graph} of about $2^{40}$ nodes.
Candidates must satisfy these three constraints:
\begin{align}
	x \text{ is candidate} \iff& \text{ } x.R3 = fixedR3 \label{eq:fixedr3}\\
	\land \quad &f(x).R3 \neq fixedR3 \label{eq:segmentleader}\\
	\land \quad &f^{-1}(x) \neq \varnothing \label{eq:nonleaf}
\end{align}
The $R3$ register of a candidate must equal an arbitrary predefined value $fixedR3$ (\ref{eq:fixedr3}).
The $R3$ register of the following state must differ from the $fixedR3$ value (\ref{eq:segmentleader}).
Leaves in the state transition graph are not considered as candidates (\ref{eq:nonleaf}).
An edge $(x,x')$ in the candidate graph corresponds to the unique path from candidate $x$ to candidate $x'$ in the state transition graph,
and is labeled with the length of that path.
Thus, in the candidate graph, nodes have an outdegree of 1 as in the state transition graph. The distribution of indegrees however will be different
in both graphs.
The candidate graph has the property being much smaller than the state transition graph $G$, but that each cycle of $G$ is still present in the candidate graph,
because each cycle of $G$ contains at least one candidate.
In general, when this property cannot be guaranteed, a random placement of candidates can be used to achieve the property with high probability, at least for larger cycles.

Removing the outgoing edges of all candidate nodes in the state transition graph $G$ partitions $G$ into \emph{segments}.
Each segment is a tree with the root being a candidate node.
%
A \emph{shallow segment} is defined as a segment that does not exceed a specific depth,
the depth $D$ being defined as the maximum distance of any node
in the segment to the segment's root.
The selection of that depth is a tuning parameter for the second step resulting in a time/space tradeoff.
The candidate node of a shallow segment must be a leaf in the candidate graph, as long as
the specified depth $D$ is smaller than the minimum distance $L$ between candidate nodes in the original graph,
 which is the case for the A5/1 graph.
In general, the candidate of a shallow segment is a leaf in the candidate graph only with high probability.

The second step performs for each candidate node a Depth First Search (DFS) in the state transition graph
 with reverse directed edges by following $f^{-1}$ to identify and remove shallow segments.
The remaining candidate nodes represent the \emph{skeleton graph} of the state transition graph.

The third step connects the nodes of the skeleton graph with weighted edges by following $f$ and thus constructs the graph.

The fourth step cuts the leaves of the skeleton graph iteratively until only cycles remain.
A fifth step calculates the number and size of the cycles in internal memory.

Further investigation is performed on the intermediate results to recover the information discarded in the first reduction step and to analyze some other properties of the A5/1 graph.

\subsection{First step: selection of candidate nodes}

The definition of a candidate reduces the state transition graph with $2^{64}$ nodes to a candidate graph with about $2^{40}$ nodes.

Restricting observation to nodes with an arbitrary $R3$ value $fixedR3$ as defined in Equation (\ref{eq:fixedr3})
 results in $(2^{19}-1) \cdot (2^{22}-1) \approx 2^{41}$ nodes.
It is also possible to fix an arbitrary $R1$ or $R2$ value, but fixing a value of the biggest register $R3$ instead leads to a smaller set of nodes and thus to a better reduction.

The direct predecessor node of a candidate could also have a $R3$ value of $fixedR3$ but different values in both registers $R1$ and $R2$.
That implies that there exist chains of nodes with the same $R3$ value $fixedR3$.
Equation (\ref{eq:segmentleader}) defines a candidate to be the last element in the chain and thus eliminates $2^{39}$ candidates, calculated as follows:

Due to the restriction of the all-zero value for each register, the number of chains depends on the value of the clocking bit $c_3$ of the predefined $R3$ value $fixedR3$.
By definition in Equation (\ref{eq:segmentleader}) the $R3$ register of a candidate will be clocked in the next clock of the stream cipher.
Multiplying the conditional probability $\mathrm{Pr}(R3 \text{ is clocked}|c_3)$ with the fixed number of nodes \textit{SR3} having the same $R3$ value $fixedR3$ gives us the expected number of chains.
Let $c_i \in \{0,1\}$ be the value of the clocking bit of register $i$.
\begin{align}
&\mathrm{E}(\text{\#Chains}|c_3) = \mathrm{Pr}(R3 \text{ is clocked}|c_3)  \cdot \mathit{SR3} \nonumber \\
&= [1- \mathrm{Pr}(R3 \text{ is not clocked}|c_3)] \cdot \mathit{SR3} \nonumber \\
&= [1- \mathrm{Pr}(c_1 = c_2 \not= c_3|c_3)] \cdot  \mathit{SR3} \nonumber \\
&= [1- \mathrm{Pr}(c_1 \not= c_3|c_3) \cdot \mathrm{Pr}(c_2 \not= c_3|c_3)] \cdot  \mathit{SR3} \nonumber \\
&= [1-\frac{2^{18}-c_3}{2^{19}-1} \cdot \frac{2^{21}-c_3}{2^{22}-1}] \cdot (2^{19}-1)(2^{22}-1) \nonumber \\
&= (2^{19}-1)(2^{22}-1) - (2^{18}-c_3)(2^{21}-c_3)
\label{eq:chain_leaders}
\end{align}
It follows that choosing a $fixedR3$ value with $c_3=0$ results in $\mathrm{E}(\text{\#Chains}|c_3=1) - \mathrm{E}(\text{\#Chains}|c_3=0) = 2359295$ (0.00014\%) fewer chains.
Therefore Equation (\ref{eq:segmentleader}) leads to a reduction of the number of candidates by $2^{39}$ ($\approx 25\%$) for $c_3=0$
but only by $2^{39} -2359295$ for $c_3=1$.

For completeness, the expected length of a chain conditioned on $c_3$ is quite close to $\approx 1.333$ nodes for both values of $c_3$ and can be calculated as follows:
\begin{align}
\mathrm{E}(\text{chain length}|c_3) = \frac{\mathit{SR3}}{\mathrm{E}(\text{\#Chains}|c_3)} \nonumber \\
= \frac{(2^{19}-1) (2^{22}-1)}{(2^{19}-1) (2^{22}-1) - (2^{18}-c_3) (2^{21}-c_3)} \nonumber
\end{align}

Another reduction of the candidate nodes is accomplished by Equation (\ref{eq:nonleaf}).
According to Golic \cite{Golic1997} the probability of an arbitrary node having no predecessors is $\frac{3}{8}$.
Keeping the two relevant bits of the $R3$ register fixed reduces the probability to $\frac{1}{4}$.
Thus another reduction of $(2^{19}-1) \cdot (2^{22}-1) \cdot \frac{1}{4} \approx 2^{39}$ candidate nodes is achieved
 and results in a total number of approximately $2^{40}$ candidate nodes.

\subsection{Second step: backward clocking}
\label{chap:back}

The second step removes the candidate nodes rooting shallow segments from the candidate graph
 by performing a depth restricted reverse DFS starting in each of the $2^{40}$ candidate nodes,
 i.e.\ it removes leaves from the candidate graph that can be identified with restricted effort,
 to further reduce the graph size.
See Section \ref{chap:results} for experimental results of the time/space tradeoff on the maximum depth of the reverse traversal described earlier.

The state transition function $f$ of the A5/1 stream cipher can be easily reverted to $f^{-1}$, so that the predecessors nodes of a node can be computed during the traversal.
As a result of the majority clock function of the A5/1 stream cipher every node can have zero or up to four direct predecessors.
Recall that the number of predecessors of a candidate node is between one and four, due to the restriction in Equation (\ref{eq:nonleaf}).

The number of predecessors depends on the clock bit and on the next higher bit of each register.
The 64 possible combinations of these six bits are covered by six rules introduced by Golic \cite{Golic1997}
 and allow to precompute look-up tables.

The $2^{40}$ DFS traversals from the candidate nodes can be executed in parallel on multiple processors
 since the traversals are performed on different segments and thus on disjoint sets of nodes.
Identifying the shallow segments and removing the associated candidate nodes
 results in a subset of the candidate nodes that represents the nodes of the skeleton graph.

Let $N$ be the number of candidates, i.e.\ the number of nodes with properties (1) to (3).
The complexity of step 2 is $O(N\cdot D)$,
 as the average number of predecessors for nodes in the state transition graph is 1,
 and thus the number of nodes expected in each level is a constant.
The step can be perfectly parallelized.

\subsection{Third step: forward clocking}
\label{chap:forward}

In this forward clocking step the nodes of the skeleton graph get connected by weighted edges.
The weight of an edge is the number of clocks of the A5/1 stream cipher needed to reach the destination node from the source node.
The algorithm in Figure \ref{algo:forward} clocks each node repeatedly until the next candidate node is reached.

The average number of clocks of the A5/1 stream cipher $L = 11184809.\overline{3}$ needed to reach a state with the same $R3$ was calculated in Section \ref{chap:basics}.
Depending on the search depth of the second step, there might be still a lot of nodes left that needs to be clocked on average $L$ times each.

The iterations of the loop can be parallelized as they are independent.
The huge number of evaluations of the state transition function requires an efficient implementation to finish this task in a reasonable time.
Therefore a GPGPU\footnote{General Purpose computing on Graphics Processing Units} bitslice approach that was used in \cite{Nohl} was adopted and executed on several GPUs in parallel.

Let $N'$ be the number of nodes in the skeleton graph resulting from step 2.
As skeleton nodes have distance $L$ from each other on average, the complexity of step 3 is $O(N'\cdot L)$.
It can be perfectly parallelized.
If step 2 would have been skipped, step 3 would take $O(N\cdot L)$.
Thus, in step 2 we must choose $D$ such that $O(N\cdot D+N'\cdot L)<O(N\cdot L)$.
This can be done by sampling a small number of segments.

\begin{figure}[ht]
\begin{minipage}[t]{0.48\linewidth}
\begin{algorithmic}
\small
\State \textbf{input} list $A$ of skeleton nodes
\State
\State $B = \varnothing$
\State  $fixedR3 \gets A[0].R3$
\ForAll{$a \in A$}
	\State $b.source \gets a$
	\State $count \gets 0$
	\Repeat
		\State $a \gets$ \textbf{clock}($a$)
		\State $count \gets count + 1$
	\Until{$a.R3 = fixedR3$}
	\While{$a$ is not candidate}
		\State $a \gets$ \textbf{clock}($a$)
		\State $count \gets count +1$
	\EndWhile
	\State $b.distance \gets count$
	\State $b.destination \gets a$
	\State $B = B \cup \{b\}$
\EndFor
\State
\State \textbf{output} list $B$ of weighted edges
\end{algorithmic}
\caption{Forward clocking}
\label{algo:forward}
\end{minipage}
\hspace{0.02\linewidth}
\begin{minipage}[t]{0.48\linewidth}
\begin{algorithmic}
\small
\State \textbf{input} list $B$ of edges with entries ($source$, $distance$, $destination$, $size$, $depth$)
\State
\ForAll{$b \in B$}
	\If{$\nexists x \in B$:  $x.destination = b.source$}
		\State $B = B \setminus \{b\}$
		\State \textbf{find}$(y \in B: y.source = b.destination)$
		\State $y.size \gets y.size + b.size + 1$
		\State $y.depth \gets $ \textbf{max}$(y.depth,\, b.depth)$
	\EndIf
\EndFor
\State
\State \textbf{output} reduced list $B$
\end{algorithmic}
\caption{Cutting leaves}
\label{alg:Cutting_leaves}
\end{minipage}
\end{figure}

\subsection{Fourth step: cutting leaves}

The resulting graph of the previous algorithm that is given as a list of edges
 is now going to be further reduced by an iterative leaf cutting algorithm, cf.\ Figure~\ref{alg:Cutting_leaves}.
Repeating this algorithm until no further reduction is gained will result in a graph containing only cycles.
The sub-tree size and depth information is preserved and stored in the root of the removed sub-tree.
A list $B$ with entries \textit{(source node, distance, destination node, sub-tree size, sub-tree depth)} is expected as input, where \textit{sub-tree size} and \textit{sub-tree depth} are initially set to zero.
Note that the $source$ entries in the input list $B$ are unique, since each node has only one outgoing edge.
The input list $B$ of edges is expected to be huge and won't fit into the internal memory of a single PC.
An efficient external memory algorithm is presented that performs only sequential operations on the list.
The parallel streaming pipeline of the STXXL\footnote{Standard Template Library for Extra Large Data Sets,
 \url{http://stxxl.sourceforge.net}} was used to implement this task.
See further details on the implementation in Section \ref{chap:impl_cutting}.

\subsection{Fifth step: count cycles}

To count the cycles and to determine their sizes an internal algorithm,
 described in Figure \ref{alg:counting},
 is applied on the remaining list of edges from the previous step.
The expected number of nodes lying on a cycle should be small enough to fit in the internal memory of a single PC.
Otherwise consider applying an efficient external cycle structure algorithm presented in \cite{KellerS01}.

\begin{figure}[h]
\begin{minipage}[t]{0.48\linewidth}
\begin{algorithmic}
\small
\State \textbf{input} edge list $B$ containing only cycles
\State
\State $visited \gets \varnothing$
\ForAll{$b \in  B$}
	\State $cycleSize \gets 0$
	\State $leader \gets \infty$
	\State \Call{findCycle}{b}
	\If {$cycleSize \neq 0$}
		\State \textbf{output} $leader,\, cycleSize$
	\EndIf
\EndFor
\end{algorithmic}
\end{minipage}
\hspace{0.02\linewidth}
\begin{minipage}[t]{0.48\linewidth}
\begin{algorithmic}
\small
\Procedure{findCycle}{$b$}
	\While {$b.source \notin visited$}
		\State $visited \gets visited \cup \{ b.source \}$
		\State $leader \gets $\textbf{min}$(leader,\, b.source)$
		\State $cycleSize \gets cycleSize + 1$
		\State \textbf{find}$(c \in B: c.source = b.destination)$
		\State $b \gets c$
	\EndWhile
\EndProcedure
\end{algorithmic}
\end{minipage}
\caption{Counting cycle lengths}
\label{alg:counting}
\end{figure}

\section{Details on the implementation}
\label{chap:impl}

\subsection{Second step: backward clocking}
The back clocking step is a crucial part of the whole task of finding the cycles in the A5/1 graph
 because the problem size of the following steps depends on the reduction factor in this step.
On each of the approximately $2^{40}$ candidate nodes an efficient limited reverse traversal must be performed.
Hybrid parallelism is applied for this task by utilizing several compute nodes of a HPC
 using MPI\footnote{Message Passing Interface} (for communication between compute nodes)
 and OpenMP\footnote{\url{http://openmp.org}} (for fully utilizing a compute node from a single process).

The main problem in this step is to generate the predecessors of a node in an efficient way.
Therefore a small (256 bytes) look-up table is precomputed
 and used to quickly identify which registers need to be clocked back
 to receive all valid predecessor nodes of the current node.
The look-up table is indexed by a six bit number
 that is composed of the concatenation of the six relevant bits (clock taps and neighbor bits) of the three registers.
For $fixedR3$ we used 0x2AAA00 ($010101010101\mathbf{0}1000000000$ in binary representation).

Another optimization is applied to the calculation of the feedback bit of the registers with more than two feedback taps ($R1$ and $R3$).
XORing several bits of a single 32-bit value can be realized by a single multiplication with a well chosen constant value.

A first node-limited (100000 nodes) BFS implementation was executed on a small cluster with 128 CPU cores.
It took 92 hours to check all candidate nodes and with the chosen tuning parameter 98.77\% of the candidate nodes were identified representing a shallow segment and thus removed.
The same configuration was executed on the LOEWE-CSC\footnote{\url{http://csc.uni-frankfurt.de/index.php?id=51}} supercomputer utilizing 2048~cores.
It took 12 hours to finish this task which is almost twice the expected time.
This loss of efficiency arises from the static way of distributing the candidate nodes to the compute nodes.
A perfect dynamic distribution would result in an expected total running time of about 6.5 hours, and will be implemented in a next version.

A depth-limited (3000 levels) DFS implementation was executed on the 128~CPU core cluster.
It takes the same amount of time as the BFS implementation but leads to a better reduction of 98.96\% shallow segments found.

\subsection{Third step: forward clocking}
\label{chap:impl_forward}

Forward clocking of each node of the skeleton graph is a huge task that requires around
 $2^{57}$
 evaluations of the clocking function of the A5/1 stream cipher.
Because of the simplicity of this algorithm and its SIMD (Single Instruction Multiple Data) nature
 it is preferred to implement and to run that job on massive parallel special purpose hardware like GPUs.

Simplified, a CUDA-enabled GPU consists of multiple Streaming Multiprocessors (SM),
 each possessing thousands of registers, a small private cache
 and a SIMD processing unit that can process 32~threads simultaneously.
All SMs have access to a large global memory.
A CUDA application runs thousands of threads, partitioned into blocks that are mapped onto the SMs.
Each block is then executed in groups of 32~threads, called warps.
Due to the SIMD nature of the processing unit, each thread of a warp has to execute the same instruction (otherwise they are sequentialized).
However, this restriction does not apply to different warps of the same block.
Intra-block communication and synchronization are fast and easily implemented using the SM's private cache,
 but inter-block communication has to be realized via the global memory and is very expensive.
For more details on architecture and programming of these GPUs we refer to the CUDA~C Programming Guide~\cite{CUDA-C-Programming-Guide}.

A first straightforward GPU implementation executed on a single NVIDIA Tesla C1060 GPU with 1.3 GHz clock frequency
 and was 85 times faster compared to a CPU implementation running on a single core with a clock frequency of 2.6 GHz.
A different concept of clocking the A5/1-registers on the GPU is introduced to speed up the computation by another factor of 4.5.
This speedup is achieved by implementing a bitslice approach \cite{Biham1997} which has previously been adopted by the A5/1 Security Project~\cite{Nohl}.

Applying the bitslice approach requires an implementation of the state transition function using only bit operations like AND, OR, XOR, and  NOT.
The input data needs to be transposed, i.e.\ an array of 32 variables each containing a 64-bit state
 needs to be transformed into an array of 64 32-bit variables where variable $v_i$ consists of the `slice' of the $i$-th bit from all input values.
Thereafter 32~evaluations of the function can be executed in parallel.
The overhead of transposing the data in internal memory before and after the GPU calculation phase must be added to the total number of operations of the bitslice method.

A straightforward implementation of the bitslice approach by having each thread of a block clocking 32~nodes in parallel results in a bad performance due to the high amount of local data per thread.
Therefore another approach is applied where each of the three warps of a single block operates on another register of the nodes.
The 64 variables that store the 32~nodes are spread among the three warps (96~threads) of a block as follows:
\begin{itemize}
\item
Thread 0-31 operates only on the R1 register consuming 19 variables of the data
\item
Thread 32-63 operates only on the R2 register consuming 22 variables of the data
\item
Thread 64-95 operates only on the R3 register consuming 23 variables of the data
\end{itemize}
The whole block of 96 threads operates on $32 \cdot 32 = 1024$ nodes in parallel.
Intra-block sync operations guarantee that the critical part of exchanging the clocking bits between the threads  through the private cache is performed in the correct order.

This algorithm is executed on NVIDIA Tesla C1060 and GTX 580 GPUs using the CUDA programming framework.
A Tesla C1060 GPU (compute capability 1.3) is able to clock 3933 nodes/s each 11170000 times
 while the newer architecture GTX 580 (compute capability 2.0) is able to clock 9800 nodes/s.
The speedup of NVIDIA's 2.0 generation mainly results from the higher amount of registers and the decrease of the warp allocation granularity, increasing the occupancy of the GPU from 38\% (compute capability 1.3) to 50\% on the newer architecture.

Due to performance reasons the GPU clocks each node a constant number of times
 (without checking whether a new candidate is reached)
 and the CPU finishes the clocking of each node while checking after each clock if the next candidate is reached.
A number of 11170000 clocks is found to be safe for the particular predefined $fixedR3$ value used but might be to high for different  $fixedR3$ values.

It takes nine days and three hours on five Tesla C1060 GPUs to finish the whole computation.
Five GTX 580 GPUs would take 4 days and 8 hours to finish the whole computation.
The reason why this implementation does not scale perfectly is because  the CPU phase and GPU phase are not executed concurrently until now, but it will be implemented in a next version.

The resulting skeleton graph contains about $2^{33.6}$ nodes and consumes 252~GB of disk space.

\subsection{Fourth step: removing leaves}
\label{chap:impl_cutting}
The edge list generated in the previous step is too big to fit in the internal memory of a single PC.
Therefore an external memory approach was implemented using the STXXL minimizing the number of I/Os.
An introduction to the STXXL library is presented in \cite{STXXL08}.
Details of external-memory algorithms are described in \cite{Vitter08}.

STXXL provides a streaming and pipelining feature so that the intermediate results do not need to be stored in temporary files,
 instead a pipeline can be implemented where each element of data is pushed through.
To speedup the internal sorting and to overlap I/O and computation the parallel pipeline feature of the STXXL is used that is described in \cite{ParallelStxxlIPDPS2009}.

The edge list  is sorted ascending according to the source node.
The source nodes $S$ are unique and the destination nodes $D$ represent a subset of the source nodes, $D \subseteq S$.
In each iteration all source nodes $S \setminus D$ are identified as leaves and removed.

\begin{figure}[p]
\begin{minipage}[b]{0.48\linewidth}
  \centering
  \includegraphics[height=0.91\textheight]{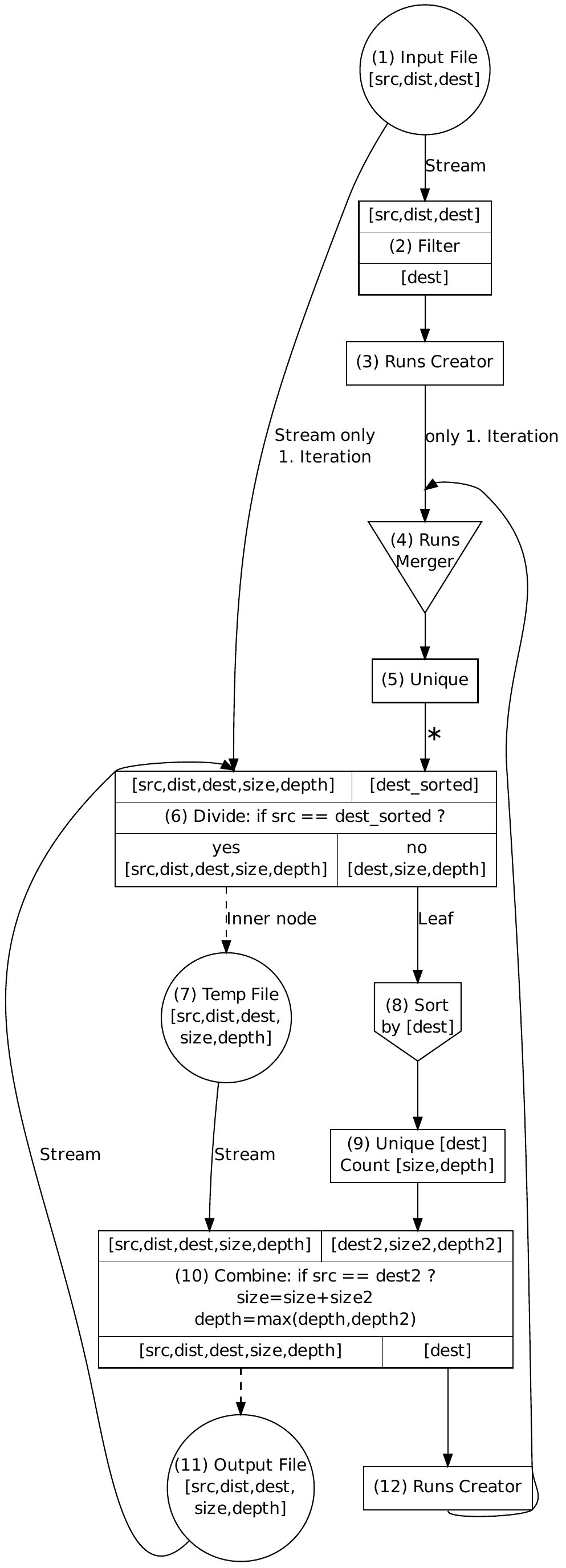}
  \caption{STXXL pipeline}
  \label{fig:pipeline}
\end{minipage}
\begin{minipage}[b]{0.48\linewidth}
\begin{center}
\begin{tabular}{|c|c|}
\hline
cycle size & number of cycles\\
\hline
1	 & 	78745\\
\hline
2	 & 	17957\\
\hline
3	 & 	7214\\
\hline
4	 & 	3647\\
\hline
5	 & 	2138\\
\hline
6	 & 	1300\\
\hline
7	 & 	867\\
\hline
8	 & 	566\\
\hline
9	 & 	441\\
\hline
10	 & 	261\\
\hline
11	 & 	181\\
\hline
12	 & 	155\\
\hline
13	 & 	105\\
\hline
14	 & 	77\\
\hline
15	 & 	66\\
\hline
16	 & 	54\\
\hline
17	 & 	50\\
\hline
18	 & 	36\\
\hline
19	 & 	19\\
\hline
20	 & 	17\\
\hline
21	 & 	12\\
\hline
22	 & 	9\\
\hline
23	 & 	6\\
\hline
24	 & 	9\\
\hline
25	 & 	5\\
\hline
26	 & 	1\\
\hline
27	 & 	2\\
\hline
28	 & 	2\\
\hline
29	 & 	4\\
\hline
31	 & 	2\\
\hline
32	 & 	1\\
\hline
33	 & 	1\\
\hline
34	 & 	5\\
\hline
38	 & 	1\\
\hline
42	 & 	1\\
\hline
\hline
total    &	113957\\
\hline
\end{tabular}
\caption{Number of cycles of the A5/1 graph with same size (cycle sizes in multiples of $L$).}
\label{tab:cycles}
\end{center}
\end{minipage}
\end{figure}

The implemented iterative parallel pipeline is shown in Figure \ref{fig:pipeline}.
The destination nodes of the edge list (1) are extracted in (2) and sorted ascending by (3) and (4).
Stage (5) generates a unique destination node stream.
At this point (*) in each iteration we can calculate the number of leaves that will be removed in the current iteration
 because the number of unique destination nodes equals the number of inner nodes.
If the number of leaves is small due to the structure of the graph, we can decide to apply another external memory heuristic to analyze the structure of the cycles.

Stage (6) performs a parallel scan on both input streams and identifies each source node as leaf if this node does not appear in the unique destination node stream.
Inner edges are pushed into a temporary file~(7) and the destination node with size and depth information of a leaf are piped to (8).

The rest of the pipeline takes care that the successor node of each leaf that was removed in this iteration
 receives the information about the number of its removed predecessors.
Therefore the destination nodes, size and depth of the removed edges in (6) are sorted in (8) ascending by the destination.
Stage (9) removes double destination nodes while summing up the size and keeping the maximum depth of each element in a run of equal destination nodes.

In (10) the size and depth information of the removed leaves is attached to their according successor node that appears in the  left input stream generated from (7).
To prepare the next iteration all edges are pushed into (11)
 and sorted runs of the destination nodes are created in (12).
In the next iteration of the loop stage (4) reads the output from (12) and (6) gets the edges streamed from (11).
The loop terminates if the number of destination nodes in the stream in (*) equals the number of edges stored in (11).

If only cycles are left the graph is small enough to perform the last step of identifying and counting the size of each cycle in internal memory as described in Figure \ref{alg:counting}.
If the resulting graph would be still large and not fit in the internal memory of a single PC consider an efficient external cycle structure algorithm presented in \cite{KellerS01}.

The computation time of the implementation is I/O bound, that means that its execution speed mainly depends on the performance of the underlying I/O system.
Experiments with the data stored on a 4 hard disk RAID-0 system showed a total running time of 48 minutes while I/O was ongoing for 93\,\% of this time with an average throughput of 350 MB/s.
The total running time of the implementation given the data stored on a 4 solid-state drive RAID-0 system was 40 minutes and I/O was active 70\,\% of the time with an average rate of 565 MB/s.

The implementation performs 88 iterations on the data which is also the maximum depth of a tree attached to a cycle in the skeleton graph.
During the whole computation a factor of 2.8 times the size of the input data was read from disk and a factor of 0.8 was written to disk.


\begin{figure}
\begin{minipage}[b]{0.48\linewidth}
  \centering
  \includegraphics[width=0.9\linewidth]{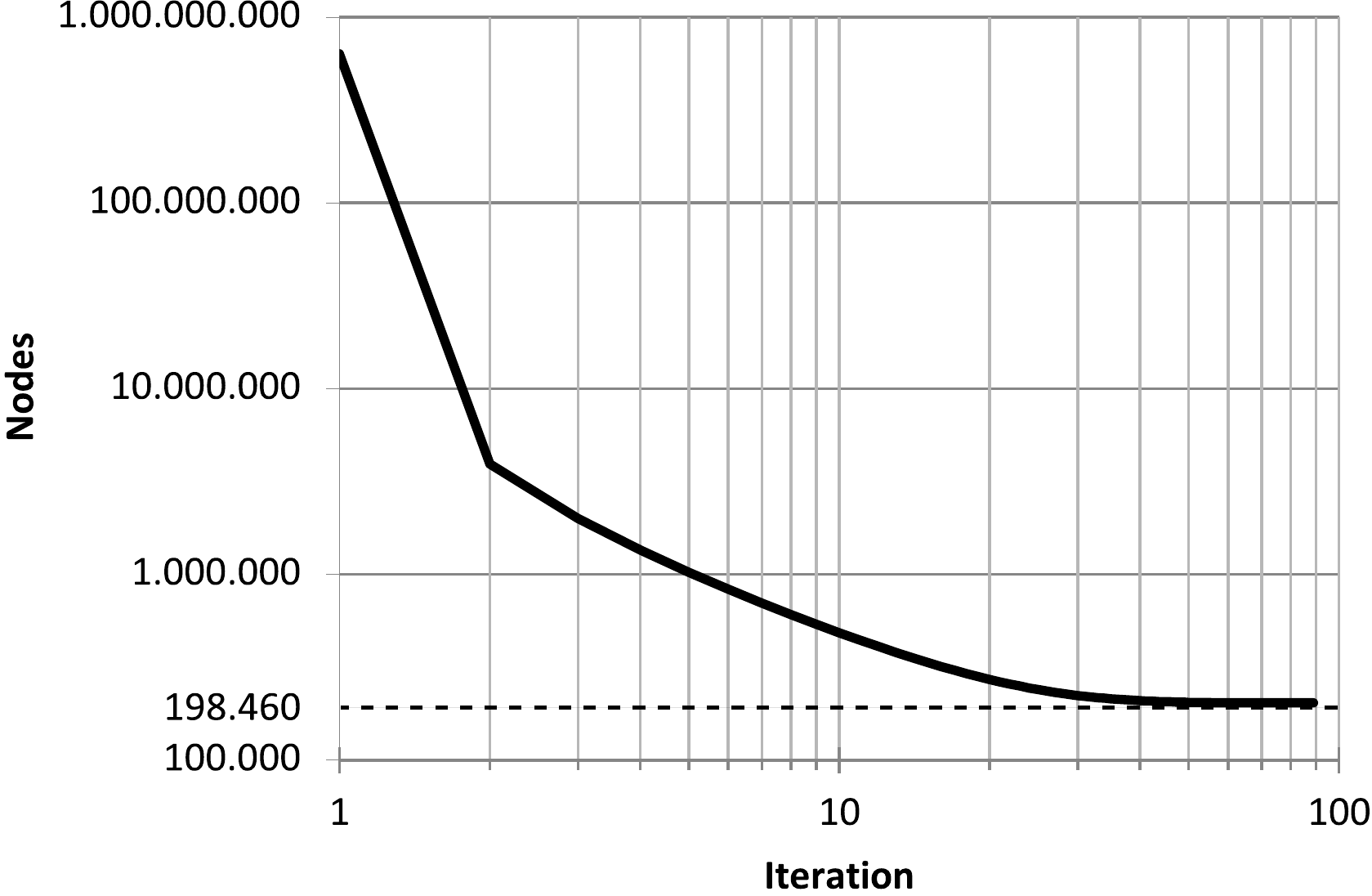}
  \caption{Reduction of nodes during the 88 iterations of the leaf removing algorithm}
  \label{fig:node_reduction}
\end{minipage}
\hspace{0.02\linewidth}
\begin{minipage}[b]{0.48\linewidth}
   \centering
   \includegraphics[width=0.8\linewidth]{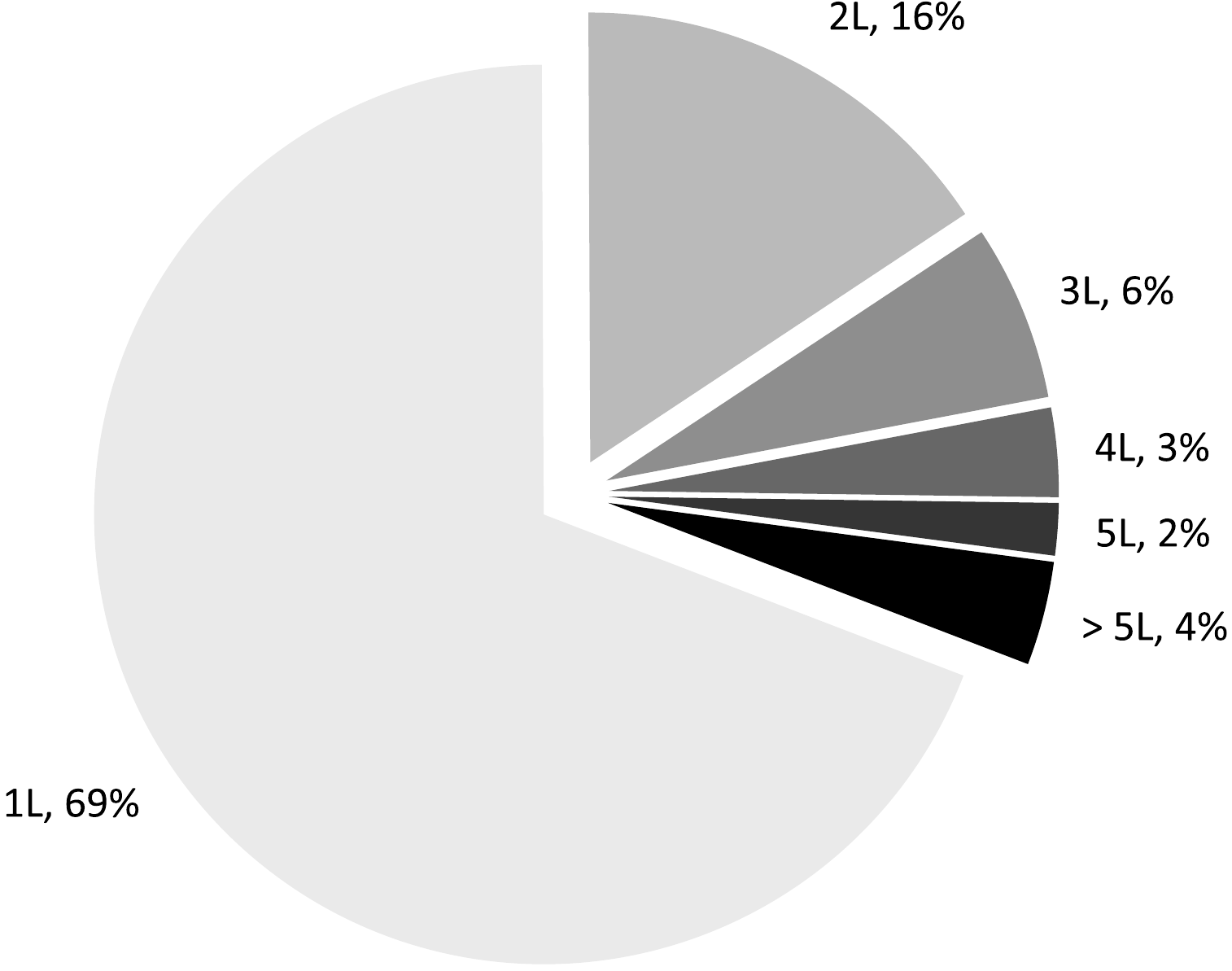}
   \caption{Distribution of the cycle sizes (indicated by $x$L) in the A5/1 graph}
   \label{fig:cycle_dist}
   \end{minipage}
\end{figure}

\section{Results}
\label{chap:results}

\subsection{Reduction of the skeleton graph}
The reduction of the nodes of the skeleton graph during the leaf cutting algorithm of the fourth step takes 88 iterations of the loop until only cycles are left.
The reduction in each iteration is presented in Figure~\ref{fig:node_reduction}.
More than 99.38\% of the nodes were identified as leaves in the first iteration.

\subsection{Number and size of cycles and components}
The total number of cycles is 113957.
The size of a cycle is a multiple of the average expected minimum length $L$ introduced in Chapter \ref{chap:basics}.
Figure~\ref{tab:cycles} lists all cycles with their specific cycle size as multiples of $L$.
Figure~\ref{fig:cycle_dist} shows an overview of the cycle size distribution in total.
It turns out that more than two thirds of the cycles have a length of just one $L$.
Traversing along the cycles and counting the nodes revealed that $2,219,735,820,460 \approx 2^{41}$ nodes
of the state transition graph's $2^{64}$ nodes
are lying on a cycle.

\subsection{Distribution of the minimum cycle length $L$}
\begin{figure}[htb]
\begin{minipage}[b]{0.48\linewidth}
\centering
\includegraphics[width=0.9\linewidth]{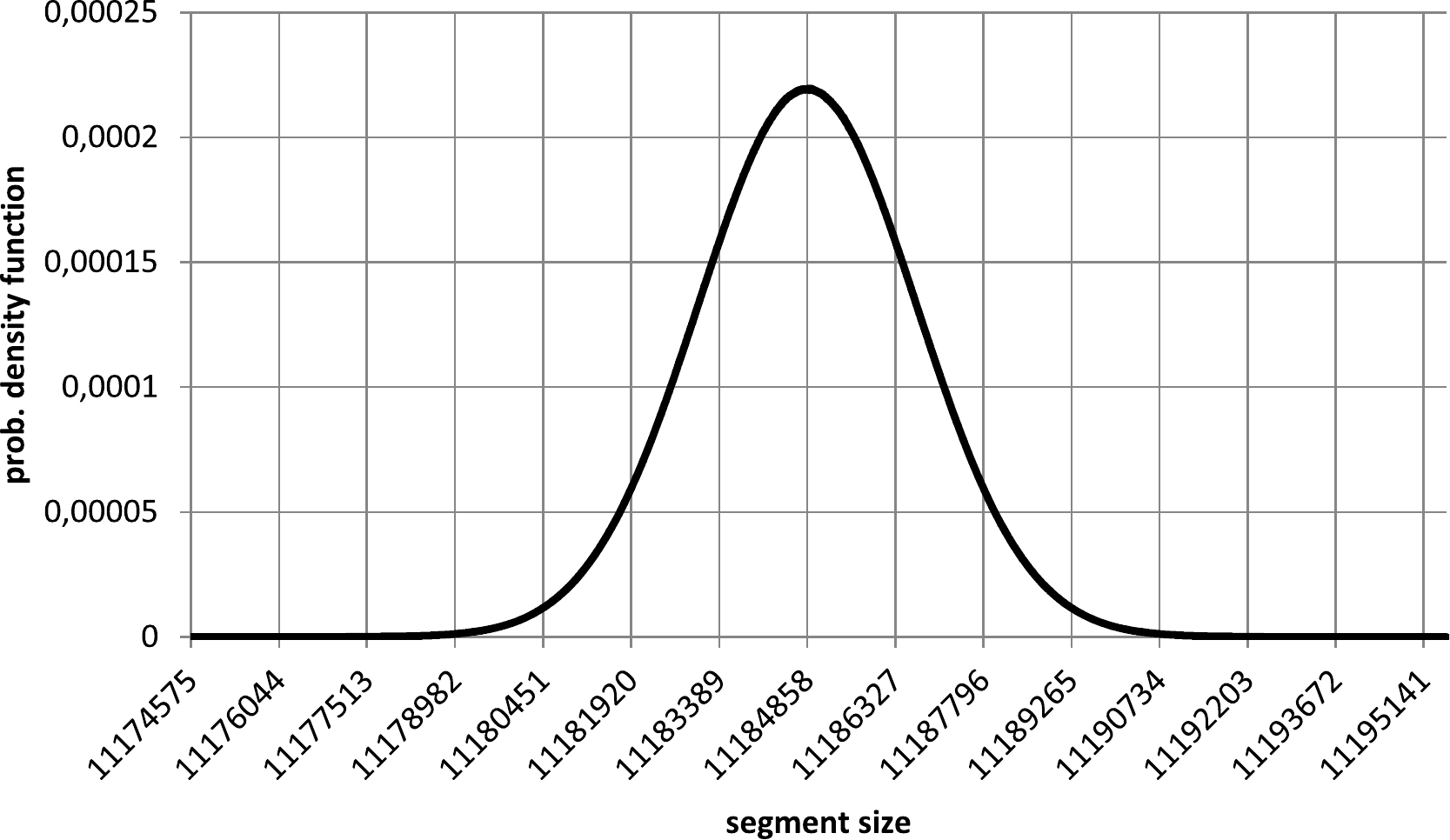}
\caption{The estimated probability distribution of the minimum cycle length $L$}
\label{fig:dist_distances}
\end{minipage}
\hspace{0.02\linewidth}
\begin{minipage}[b]{0.48\linewidth}
\centering
\includegraphics[width=0.9\linewidth]{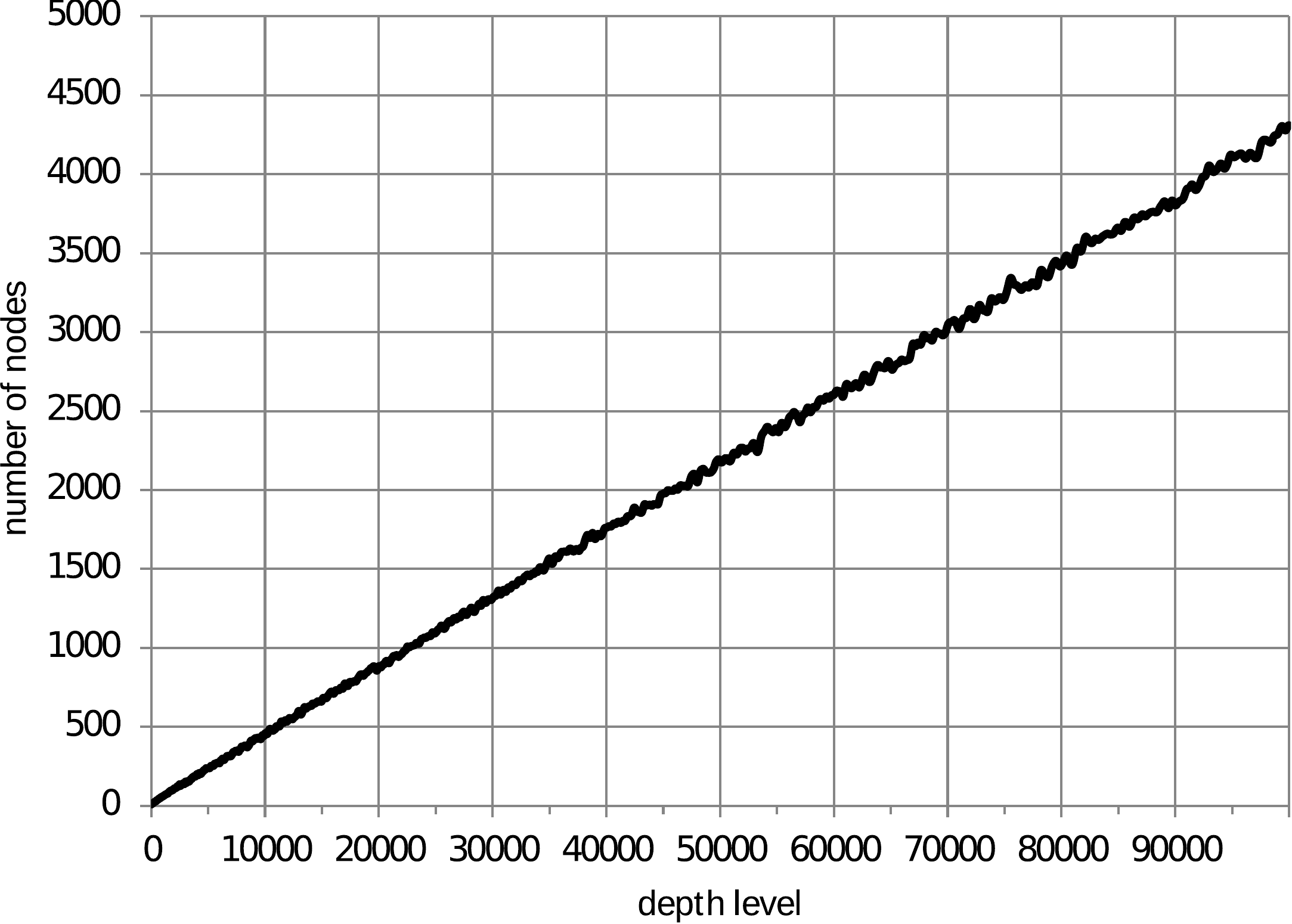}
\caption{Expected number of nodes on a specific depth level}
\label{fig:nodes_on_level}
\end{minipage}
\end{figure}

Since the minimum cycle length $L$ deviates around the mean value of $11184809\frac{1}{3}$ it is interesting to see how the real minimum cycle lengths are distributed around the mean value.
Therefore a random set of around $2^{33}$ nodes was generated and analyzed.
From the sample set it turns out that the experimental expected segment size $\mu = 11184857$ is quite close to the expected $L$.
The standard deviation of the sample set is around $\sigma = 1818$ leading to normal distribution plotted in Figure \ref{fig:dist_distances}.

\subsection{Number of nodes on depth level}
The following data is generated by traversing the segments of a random sample set of $2^{26}$ nodes keeping the $R3$ register constant.
For each node of the random sample set the number of nodes in each depth level was counted,
 the depth of a node being its distance from segment's root, i.e.\ candidate, in the state transition graph $G$.
The average number of nodes on a specific depth level is shown in  Figure \ref{fig:nodes_on_level}.
Note that the graph in Figure \ref{fig:nodes_on_level} does not take into account the probability of a traversal  reaching that depth, instead it gives the average number of nodes expected to be found on a specific level if a traversal reaches that depth.
Taking the probability of reaching that depth into account the number of expected nodes on a specific depth level is a constant of $\approx 1.7$.

\subsection{Segment depths}
\begin{figure}
\begin{minipage}{0.48\linewidth}
	\centering
	\includegraphics[width=0.9\textwidth]{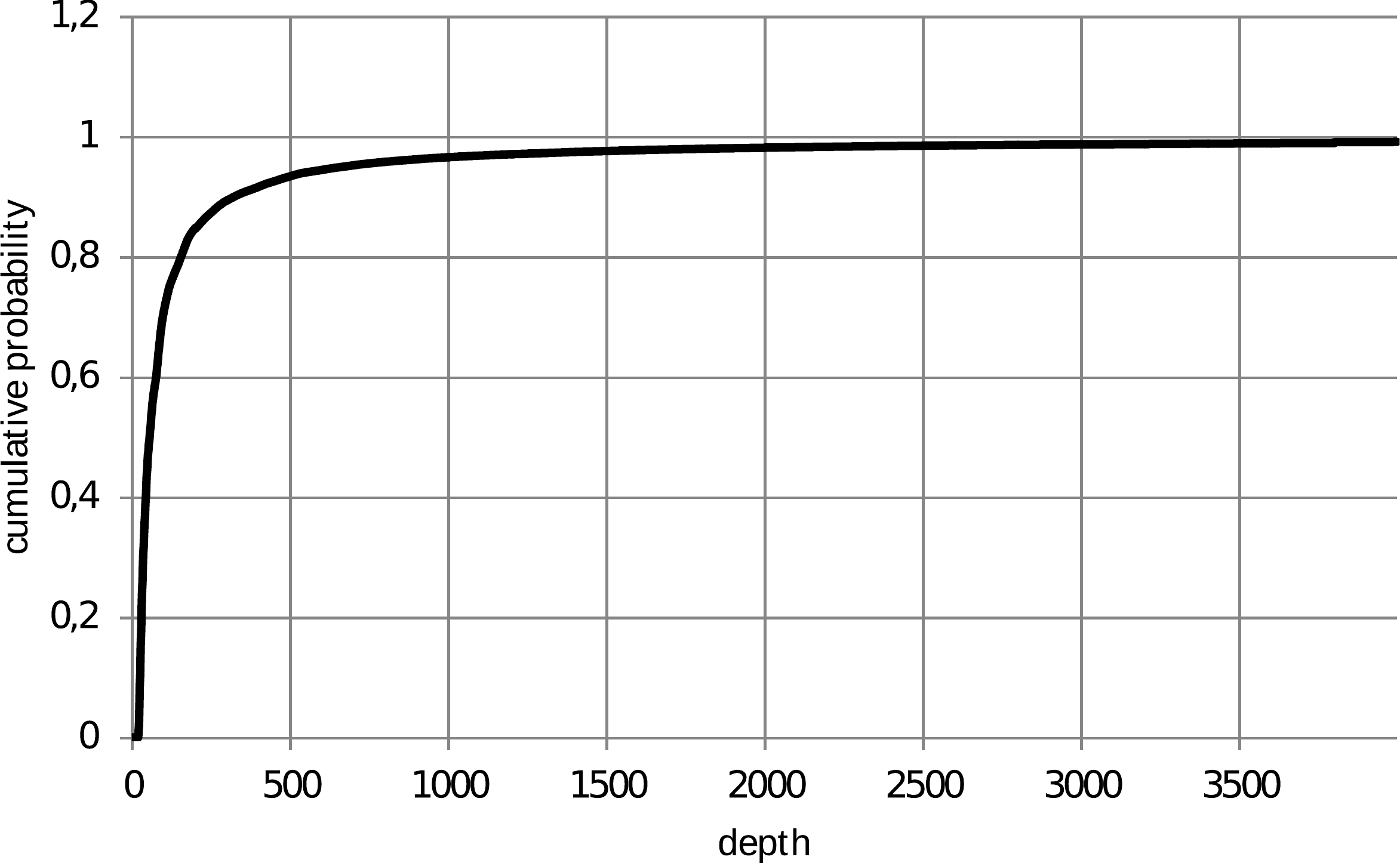}
	\caption{The cumulative distribution function of the segment depth}
	\label{fig:subtree_depth}
\end{minipage}
\hspace{0.02\linewidth}
\begin{minipage}{0.48\linewidth}
\begin{center}
\begin{tabular}{|c|c|}
\hline
fraction of segments & maximum depth \\
\hline
50\% & 56\\
90\% & 319\\
99\% & 3539\\
99.9\% & 39065\\
\hline
\end{tabular}
\end{center}
\caption{Maximum depth of segments}
\label{tab:max_depth}
\end{minipage}
\end{figure}

Another interesting fact is the expected maximum depth of a segment.
There is a time/space tradeoff on the maximum search depth to be chosen in the backward clocking step.
Therefore the maximum segment depth of a random sample set of around $2^{26}$ nodes is calculated and the cumulative distribution function of the result is plotted in Figure \ref{fig:subtree_depth}.
Note that the $R3$ register of each node in that random sample set is kept constant as in the backward clocking algorithm.
It turns out that 50\% of the segments have a depth less than 56, other prominent values are shown in Figure~\ref{tab:max_depth}.

\subsection{Results compared to expected properties of random graph}
In order to verify if the graph generated by the A5/1 stream cipher equals a graph generated by a random mapping, the results are compared to the the expected properties of random mappings in Table \ref{tab:comparison}.
The total number of nodes in the A5/1 graph is $n=(2^{19}-1)\cdot(2^{22}-1)\cdot(2^{23}-1)$.
Exact values are given if observed.

\begin{table}[b]
\begin{center}
\begin{tabular}{|c|c|c|c|c|}
\hline
property & formula \cite{Flajolet1990} & expected value & experimental value & deviation\\
\hline
\# of components: 	& $\frac{1}{2}\log n$ 		& 22			& 113957				& $10^{3.71}$ \\
largest cycle: 		& $\approx 0.78248\sqrt{n}$ 	& $\approx 2^{31.6}$	& $\approx 2^{28.81} (469758320)$	& $10^{-0.84}$ \\
largest component: 	& $\approx 0.75782n$ 		& $\approx 2^{63.6}$	& $\approx 2^{52.34}$			& $10^{-3.39}$ \\
max segment depth: 	& $0.6 \sqrt{n}$ 		& $\approx 2^{31.3}$	& $\approx 2^{29.87}$ 			& $10^{-0.43}$ \\
fraction of leaves:	& $1/e$				& $\approx 36.79\%$	& \small$\approx 37.5\% (6917518582283894784)$& $10^{0.01}$ \\
\# of cycle nodes:	& $\sqrt{\pi\frac{n}{2}}$ 	& $\approx 2^{32.3}$	& $\approx 2^{41} (2219735820460)$ 	& $10^{2.62}$ \\
\# of nodes on level:	& 1	 			& 1 			& $\approx 1.7$ 			& $10^{0.23}$ \\
\hline
\end{tabular}
\end{center}
\caption{Comparison of the expected values to the results}
\label{tab:comparison}
\end{table}

The experimental results lead to the conclusion that the structure of the A5/1 graph is non-random.

\begin{figure}[t]
	\centering
	\includegraphics[scale=0.3, angle=-90]{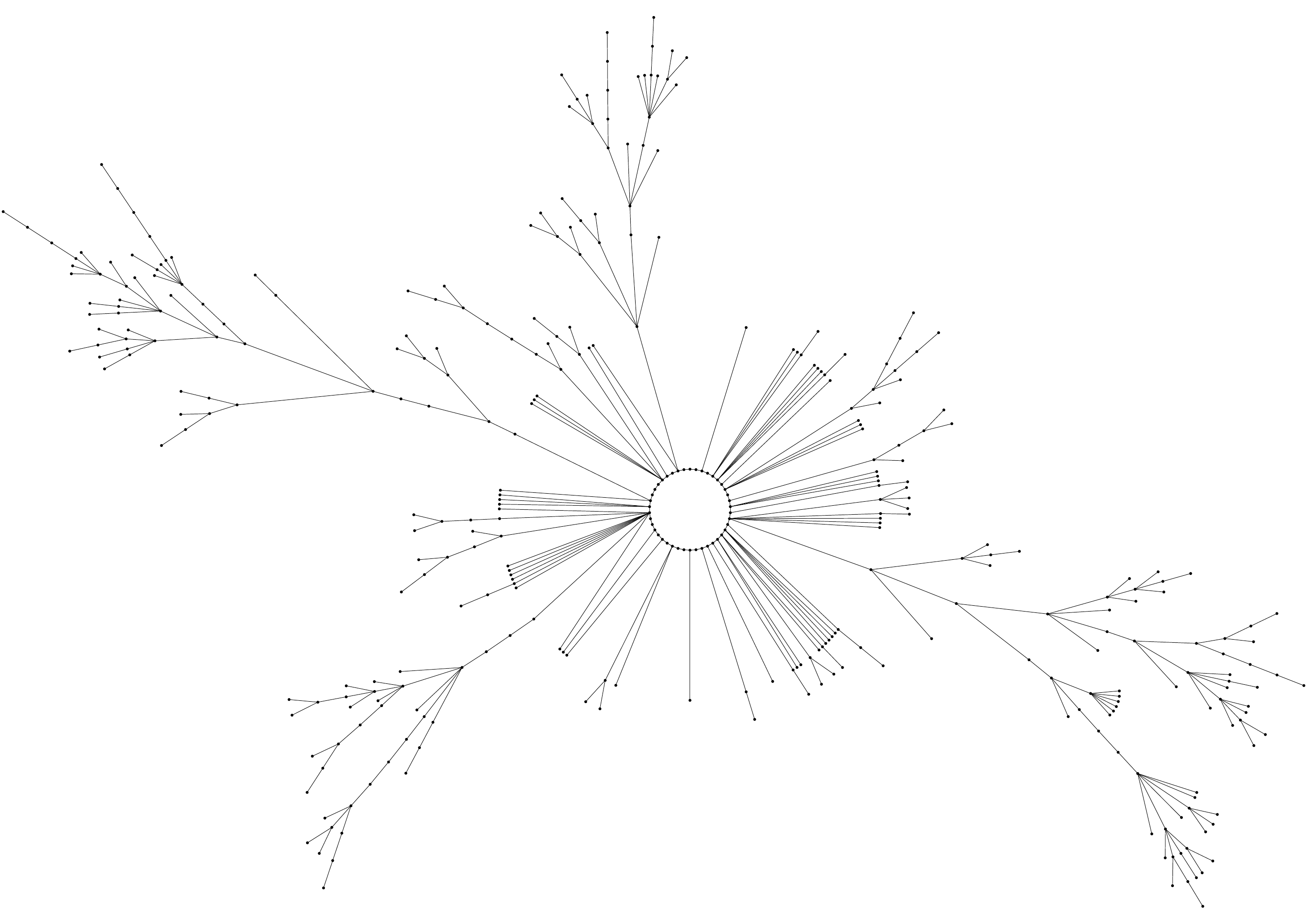}
	\caption{The skeleton graph of the WCC with the largest cycle of 42 segments after one iteration of leaf cutting}
	\label{fig:cycle_42}
\end{figure}

\subsection{Size and structure of WCCs}
The size of the skeleton graph of a WCC varies between four and 4194575 segments before applying any leaf cutting iteration.
After one iteration of leaf cutting the sizes varies between 1 and 1210 segments.
The skeleton graph of the WCC with the largest cycle of $42 \cdot L$ and a total number of 370 segments after one iteration of leaf cutting is presented in Figure \ref{fig:cycle_42}.

\section{Conclusion and further work}
\label{chap:conclusion}
The structure of the graph induced by the A5/1 stream cipher is completely analyzed using state-of-the-art techniques and methods.
Most of the results presented in the previous chapter differ from the expected values, thus the structure of this graph notably deviates from a random mapping.

A five step approach is applied to retrieve these results.
Each step is optimized by hand for best performance considering the special aspects and properties of the graph produced by the A5/1 stream cipher.
With these optimizations the properties of the graph can be calculated in a reasonable time.

Massive parallel computations on the GPU is an interesting field with high potentials in applications on huge data sets.
Traditional algorithms must be ported under consideration of the different architecture of this special hardware in order to make use of the high computing performance of modern GPUs.
New hardware versions and architectures of GPUs are provided by the major manufacturers to satisfy the computing needs of modern applications.

Most of the algorithms presented in this work are optimized for the A5/1 graph but some ideas can be ported to other graphs produced by other state transition functions.
The scaling behavior of the implementations of the algorithms can be further improved with the ideas mentioned in chapter \ref{chap:impl}.

\bibliographystyle{eptcs}
\bibliography{literatur}

\begin{thebibliography}{10}
\providecommand{\bibitemdeclare}[2]{}
\providecommand{\surnamestart}{}
\providecommand{\surnameend}{}
\providecommand{\urlprefix}{Available at }
\providecommand{\url}[1]{\texttt{#1}}
\providecommand{\href}[2]{\texttt{#2}}
\providecommand{\urlalt}[2]{\href{#1}{#2}}
\providecommand{\doi}[1]{doi:\urlalt{http://dx.doi.org/#1}{#1}}
\providecommand{\bibinfo}[2]{#2}

\bibitemdeclare{inproceedings}{ParallelStxxlIPDPS2009}
\bibitem{ParallelStxxlIPDPS2009}
\bibinfo{author}{Andreas \surnamestart Beckmann\surnameend},
  \bibinfo{author}{Roman \surnamestart Dementiev\surnameend} \&
  \bibinfo{author}{Johannes \surnamestart Singler\surnameend}
  (\bibinfo{year}{2009}): \emph{\bibinfo{title}{Building a parallel pipelined
  external memory algorithm library}}.
\newblock In: {\sl \bibinfo{booktitle}{Parallel Distributed Processing -- IPDPS
  2009}}, \bibinfo{publisher}{IEEE}, pp. \bibinfo{pages}{1--10},
  \doi{10.1109/IPDPS.2009.5161001}.

\bibitemdeclare{inproceedings}{Beckmann2007}
\bibitem{Beckmann2007}
\bibinfo{author}{Andreas \surnamestart Beckmann\surnameend} \&
  \bibinfo{author}{Jörg \surnamestart Keller\surnameend}
  (\bibinfo{year}{2007}): \emph{\bibinfo{title}{Parallel-External Computation
  of the Cycle Structure of Invertible Cryptographic Functions}}.
\newblock In: {\sl \bibinfo{booktitle}{Proc. 15th EUROMICRO Int. Conf.
  Parallel, Distributed and Network-Based Processing PDP '07}}, pp.
  \bibinfo{pages}{526--533}, \doi{10.1109/PDP.2007.61}.

\bibitemdeclare{incollection}{Biham1997}
\bibitem{Biham1997}
\bibinfo{author}{Eli \surnamestart Biham\surnameend} (\bibinfo{year}{1997}):
  \emph{\bibinfo{title}{A fast new {DES} implementation in software}}.
\newblock In \bibinfo{editor}{Eli \surnamestart Biham\surnameend}, editor: {\sl
  \bibinfo{booktitle}{Fast Software Encryption}}, {\sl \bibinfo{series}{Lecture
  Notes in Computer Science}} \bibinfo{volume}{1267},
  \bibinfo{publisher}{Springer Berlin / Heidelberg}, pp.
  \bibinfo{pages}{260--272}, \doi{10.1007/BFb0052352}.

\bibitemdeclare{incollection}{Biham2000}
\bibitem{Biham2000}
\bibinfo{author}{Eli \surnamestart Biham\surnameend} \& \bibinfo{author}{Orr
  \surnamestart Dunkelman\surnameend} (\bibinfo{year}{2000}):
  \emph{\bibinfo{title}{Cryptanalysis of the {A5/1} {GSM} Stream Cipher}}.
\newblock In \bibinfo{editor}{Bimal \surnamestart Roy\surnameend} \&
  \bibinfo{editor}{Eiji \surnamestart Okamoto\surnameend}, editors: {\sl
  \bibinfo{booktitle}{Progress in Cryptology -- INDOCRYPT 2000}}, {\sl
  \bibinfo{series}{Lecture Notes in Computer Science}} \bibinfo{volume}{1977},
  \bibinfo{publisher}{Springer Berlin / Heidelberg}, pp.
  \bibinfo{pages}{43--51}, \doi{10.1007/3-540-44495-5\_5}.

\bibitemdeclare{misc}{cryptome:A5}
\bibitem{cryptome:A5}
\bibinfo{author}{\surnamestart Cryptome.org\surnameend}:
  \emph{\bibinfo{title}{GSM A5 Files Published on Cryptome and Elsewhere}}.
\newblock \urlprefix\url{http://cryptome.org/0001/gsm-a5-files.htm}.
\newblock \bibinfo{note}{(accessed January 10, 2012)}.

\bibitemdeclare{article}{STXXL08}
\bibitem{STXXL08}
\bibinfo{author}{Roman \surnamestart Dementiev\surnameend},
  \bibinfo{author}{Lutz \surnamestart Kettner\surnameend} \&
  \bibinfo{author}{Peter \surnamestart Sanders\surnameend}
  (\bibinfo{year}{2008}): \emph{\bibinfo{title}{{STXXL}: standard template
  library for {XXL} data sets}}.
\newblock {\sl \bibinfo{journal}{Software: Practice and Experience}}
  \bibinfo{volume}{38}(\bibinfo{number}{6}), pp. \bibinfo{pages}{589--637},
  \doi{10.1002/spe.844}.

\bibitemdeclare{incollection}{Flajolet1990}
\bibitem{Flajolet1990}
\bibinfo{author}{Philippe \surnamestart Flajolet\surnameend} \&
  \bibinfo{author}{Andrew \surnamestart Odlyzko\surnameend}
  (\bibinfo{year}{1990}): \emph{\bibinfo{title}{Random Mapping Statistics}}.
\newblock In \bibinfo{editor}{Jean-Jacques \surnamestart Quisquater\surnameend}
  \& \bibinfo{editor}{Joos \surnamestart Vandewalle\surnameend}, editors: {\sl
  \bibinfo{booktitle}{Advances in Cryptology -- EUROCRYPT ’89}}, {\sl
  \bibinfo{series}{Lecture Notes in Computer Science}} \bibinfo{volume}{434},
  \bibinfo{publisher}{Springer Berlin / Heidelberg}, pp.
  \bibinfo{pages}{329--354}, \doi{10.1007/3-540-46885-4\_34}.

\bibitemdeclare{incollection}{Golic1997}
\bibitem{Golic1997}
\bibinfo{author}{Jovan~Dj. \surnamestart Golić\surnameend}
  (\bibinfo{year}{1997}): \emph{\bibinfo{title}{Cryptanalysis of Alleged A5
  Stream Cipher}}.
\newblock In \bibinfo{editor}{Walter \surnamestart Fumy\surnameend}, editor:
  {\sl \bibinfo{booktitle}{Advances in Cryptology -- EUROCRYPT ’97}}, {\sl
  \bibinfo{series}{Lecture Notes in Computer Science}} \bibinfo{volume}{1233},
  \bibinfo{publisher}{Springer Berlin / Heidelberg}, pp.
  \bibinfo{pages}{239--255}, \doi{10.1007/3-540-69053-0\_17}.

\bibitemdeclare{misc}{GSM2010}
\bibitem{GSM2010}
\bibinfo{author}{Wireless~Intelligence \surnamestart of~GSMA
  Media~LLC\surnameend} (\bibinfo{year}{February 1, 2011}):
  \emph{\bibinfo{title}{Number of connections by bearer technology {Q4} 2010}}.
\newblock \bibinfo{note}{Received via email}.

\bibitemdeclare{incollection}{Keller2007}
\bibitem{Keller2007}
\bibinfo{author}{Jörg \surnamestart Keller\surnameend} (\bibinfo{year}{2007}):
  \emph{\bibinfo{title}{Efficient Sampling of the Structure of Crypto
  Generators’ State Transition Graphs}}.
\newblock In \bibinfo{editor}{Andrew \surnamestart Blyth\surnameend} \&
  \bibinfo{editor}{Iain \surnamestart Sutherland\surnameend}, editors: {\sl
  \bibinfo{booktitle}{EC2ND 2006}}, \bibinfo{publisher}{Springer London}, pp.
  \bibinfo{pages}{3--12}, \doi{10.1007/978-1-84628-750-3\_1}.

\bibitemdeclare{incollection}{KellerS01}
\bibitem{KellerS01}
\bibinfo{author}{Jörg \surnamestart Keller\surnameend} \& \bibinfo{author}{Jop
  \surnamestart Sibeyn\surnameend} (\bibinfo{year}{2001}):
  \emph{\bibinfo{title}{Beyond External Computing: Analysis of the Cycle
  Structure of Permutations}}.
\newblock In \bibinfo{editor}{Rizos \surnamestart Sakellariou\surnameend}
  et~al., editors: {\sl \bibinfo{booktitle}{Euro-Par 2001 Parallel
  Processing}}, {\sl \bibinfo{series}{Lecture Notes in Computer Science}}
  \bibinfo{volume}{2150}, \bibinfo{publisher}{Springer Berlin / Heidelberg},
  pp. \bibinfo{pages}{333--342}, \doi{10.1007/3-540-44681-8\_48}.

\bibitemdeclare{article}{Lamport81a}
\bibitem{Lamport81a}
\bibinfo{author}{Leslie \surnamestart Lamport\surnameend}
  (\bibinfo{year}{1981}): \emph{\bibinfo{title}{Password authentication with
  insecure communication}}.
\newblock {\sl \bibinfo{journal}{Commun. ACM}} \bibinfo{volume}{24}, pp.
  \bibinfo{pages}{770--772}, \doi{10.1145/358790.358797}.

\bibitemdeclare{misc}{Nohl}
\bibitem{Nohl}
\bibinfo{author}{Karsten \surnamestart Nohl\surnameend} \&
  \bibinfo{author}{Sascha \surnamestart Krissler\surnameend}:
  \emph{\bibinfo{title}{A5/1 Security Project}}.
\newblock \urlprefix\url{http://reflextor.com/trac/a51}.
\newblock \bibinfo{note}{(accessed January 10, 2012)}.

\bibitemdeclare{misc}{CUDA-C-Programming-Guide}
\bibitem{CUDA-C-Programming-Guide}
\bibinfo{author}{\surnamestart NVIDIA\surnameend}: \emph{\bibinfo{title}{CUDA C
  Programming Guide}}.
\newblock
  \urlprefix\url{http://developer.download.nvidia.com/compute/DevZone/docs/htm%
l/C/doc/CUDA_C_Programming_Guide.pdf}.

\bibitemdeclare{book}{Vitter08}
\bibitem{Vitter08}
\bibinfo{author}{Jeffrey~S. \surnamestart Vitter\surnameend}
  (\bibinfo{year}{2008}): \emph{\bibinfo{title}{Algorithms and Data Structures
  for External Memory}}.
\newblock \bibinfo{publisher}{now Publishers}, \doi{10.1561/0400000014}.

\end{thebibliography}

\end{document}